\documentclass[aps,showpacs,prl,twocolumn,superscriptaddress]{revtex4}
\usepackage{mathptmx}
\usepackage{mathrsfs}
\usepackage{graphicx}
\usepackage{dcolumn}
\usepackage{bm}
\usepackage{color}

\begin{document}

\title{Bragg scattering of light in vacuum structured by strong periodic fields}
\author{Gagik Yu. Kryuchkyan}
\email{kryuchkyan@ysu.am}
\affiliation{Max-Planck-Institut f\"ur Kernphysik, Saupfercheckweg 1, Heidelberg 69117, Germany}
\affiliation{Yerevan State University, Alex Manoogian Street 1, Yerevan 0025, Armenia}
\affiliation{Institute for Physical Research, National Academy of Sciences, Ashtarak 2, 0203, Armenia}

\author{Karen Z. Hatsagortsyan}
\email{k.hatsagortsyan@mpi-k.de}
	\affiliation{Max-Planck-Institut f\"ur Kernphysik, Saupfercheckweg 1, Heidelberg 69117, Germany}

\begin{abstract}

Elastic scattering of laser radiation due to vacuum polarization by  spatially modulated strong electromagnetic fields   is considered. The  Bragg interference arising at a specific impinging direction of the probe wave concentrates the scattered light in specular directions. The interference maxima are enhanced  with respect to the usual  vacuum polarization effect proportional to the square of the number of modulation periods within the interaction region. The Bragg scattering can be employed to detect the vacuum polarization effect in a setup of multiple crossed super-strong laser beams with parameters envisaged in the future Extreme Light Infrastructure.

\end{abstract}

\maketitle

Quantum electrodynamical vacuum fluctuates via virtual electron-positron pairs
which induces polarization in strong external fields.
The characteristic field when the vacuum polarization effects become significant is the
so-called critical field $E_{cr}=m^2/e$ at which an electron gains energy equal to its rest mass $m$
within the Compton wavelength $\lambdabar_c=1/m$
\cite{Sauter,Heisenberg,Schwinger}, where $e$ is the electron charge, $I_{cr}\equiv E_{cr}^2/8\pi\approx 2.3\times 10^{29}$ W/cm$^2$, $\hbar=c=1$
units are used throughout.
The only vacuum polarization effect observed experimentally is
the Delbr\"uck scattering \cite{Delbruck_exp}, the scattering of $\gamma$-rays by a high-Z atomic target.
In this process vacuum is polarized due to singular Coulomb field of highly charged ions.
In contrast to that, vacuum polarization effects  are not observed in macroscopic electromagnetic fields created in a laboratory.
Since 2000 the experiments of the PVLAS collaboration has been underway \cite{PVLAS,PVLAS1} devoted to measurement of
vacuum birefringence in a constant magnetic field
using state-of-the-art technique of superconducting magnets with a magnetic field $B$ reaching $B/B_{cr}\sim 10^{-9}$, with the critical magnetic field $B_{cr}=m^2/e=4.4\times 10^9$ T.
Recently, strong field laser technique is advancing rapidly, fostered, from one side,
by the laser fusion program \cite{NIF} and, from another side, by the Extreme Light
Infrastructure (ELI) project \cite{ELI} aiming at the creation of the strongest laser fields 
for scientific purposes 
using all the power of the chirped pulse amplification technique \cite{CPA}.
The laser field is the strongest field created in a laboratory which can be harnessed to test
the strong field QED theory via vacuum polarization effects \cite{Biryula,Itzikson,Marklund, Salamin}.
The present petawatt lasers can produce intensities of $I\sim10^{22}$  W/cm$^2$ \cite{Yanovsky},
while intensities up to $I\sim10^{26}$  W/cm$^2$ ($E/E_{cr}\sim 3\times10^{-2}$) are envisaged in the ELI \cite{ELI}.

\begin{figure}
\begin{center}
 \includegraphics[width=0.43\textwidth]{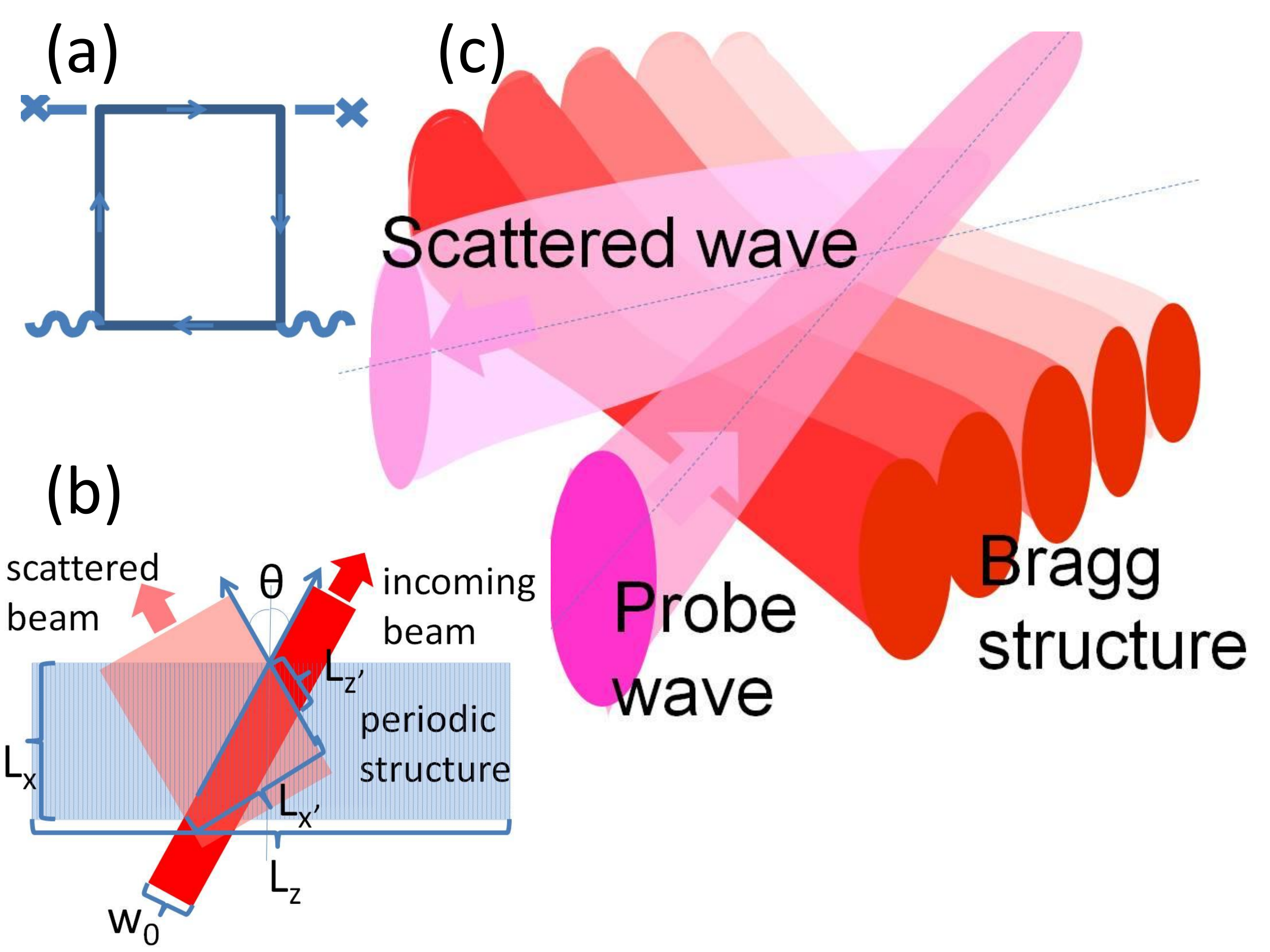}
 \caption{(Color online) (a) The Feynman diagram describing the photon scattering in a strong external electromagnetic field. (b) The geometry of the interaction region formed by overlap of the incoming laser beam with the periodic structure of the external field. (c) Bragg scattering of a probe laser beam by a set of focused strong laser beams. } \label{setup}
\end{center}
\end{figure}

Even in the strongest laser fields the vacuum polarization effects are perturbative  because of smallness of the characteristic parameter $E/E_{cr}\ll 1$. In terms of Feynman diagrams, the largest contribution to vacuum polarization arises from the box diagram, see Fig. \ref{setup}(a). 
The depicted diagram with two legs belonging to a constant uniform magnetic field or to an external laser field with a uniform amplitude describes the scattering to zero angle \cite{Remark1} that induces
the vacuum refractive index. Thus, in these cases, vacuum behaves as a uniform birefringent medium \cite{Dittrich}. Possibilities to observe vacuum birefringence in strong laser fields have been  recently discussed in detail \cite{Heinzl,Marklund}. The box diagram with one leg describing the external field does not exist in a constant uniform magnetic field \cite{Adler},
while in a laser field it describes the photon-photon scattering process \cite{LandauQED}. The attempts to observe photon-photon scattering with laser beams were  only able to determine the upper limit of the cross-section \cite{Moulin}. The possibility of observation of the photon-photon scattering with modern strong laser beams is analyzed in \cite{DiP05,Lundstrom,Stenflo}.
In the external field nonuniform in space, vacuum behaves as a nonuniform medium and photon scattering (or diffraction) becomes possible. The vacuum polarization effects due to a spatial gradient of magnetic field are considered in \cite{Kaplan}. Diffraction effects due to the vacuum nonuniform polarization in strong focused laser beams are investigated in \cite{DiP,King}. In optics of continuous media it is known that the periodic structure of a medium can significantly enhance the scattering  due to interference of the scattered light generated from different layers of the structure (Bragg scattering) \cite{Landau}.
The Bragg concept is quite general and is applied not only in the
context of propagation of electromagnetic waves \cite{Solomon} but also in quantum optics \cite{kry},
atom optics \cite{KapitzaDirac} and for matter waves \cite{Cronin}.

In this letter, we investigate Bragg scattering of a probe laser beam due to vacuum polarization in a strong spatially modulated external electromagnetic field. At certain scattering angles when the Bragg interference condition is fulfilled, the probability of the  photon scattering  is enhanced with respect to the usual photon-photon scattering by a factor proportional to the square of the number of periods in the structure. The enhancement is maintained also in the total probability of the scattering, integrated by scattering angle. First, we show the effect on the theoretically more transparent case of a spatially periodic magnetic field of a magnetic undulator. Then, we discuss the Bragg scattering when a probe laser beam penetrates the periodic structure of multiple focused laser beams, see Fig.~\ref{setup}(c).
We consider the experimental realization of the Bragg scattering in the future ELI facility and its advantage with respect to the photon-photon scattering \cite{Lundstrom}.

We consider a photon scattering in an external field, i.e., the transition $|1(\textbf{k}_1,\textbf{e}_1) \rangle \rightarrow |1(\textbf{k}_2,\textbf{e}_2) \rangle$, where  $\textbf{k}_{1,2}$ and $\textbf{e}_{1,2}$ are the  momentum and polarization  of the incoming and outgoing photons, respectively.
The transition matrix element reads:
\begin{equation}
 \label{Transition}
T_{21}=-i\langle  1(\textbf{k}_2,\textbf{e}_2)|\int_{-\infty}^{\infty}dtd^3\textbf{r}\,\,j_{\mu}(\textbf{r},t)A^{\mu}(\textbf{r},t)|1(\textbf{k}_1,\textbf{e}_1) \rangle,
\end{equation}
where $A^{\mu}(\textbf{r})$ is the field vector-potential and $j_{\mu}(\textbf{r},t)$ the current density of the polarized vacuum: $\textbf{j}=\frac{\partial \textbf{P}}{\partial t}+\bf{\nabla}\times\textbf{M}$. The vacuum polarization $\textbf{P}$ and magnetization $\textbf{M}$ are derived from the Euler-Heisenberg Lagrangian \cite{LandauQED}:
\begin{eqnarray}
\mathbf{P}(\mathbf{r},t)&=&\frac{\eta}{ 4\pi}\left[2(E^2-B^2)\mathbf{E}+7(\mathbf{E}\cdot\mathbf{B})\mathbf{B}\right], \nonumber\\
\mathbf{M}(\mathbf{r},t)&=&\frac{\eta}{4\pi}\left[-2(E^2-B^2)\mathbf{B}+7(\mathbf{E}\cdot\mathbf{B})\mathbf{E}\right],
\end{eqnarray}
where $\eta=e^4/45\pi m^4$. $\mathbf{E}(\mathbf{r},t)$ and $\mathbf{B}(\mathbf{r},t)$ are the total electric and magnetic fields, respectively, consisting of the strong external field (marked by an index $^{(0)}$) and the quantized probe field (marked by an index $^{(p)}$): $\textbf{E}(\mathbf{r},t)=\textbf{E}^{(0)}(\textbf{r},t)+\textbf{E}^{(p)}(\textbf{r},t)$, $\textbf{B}(\mathbf{r},t)=\textbf{B}^{(0)}(\textbf{r},t)+\textbf{B}^{(p)}(\textbf{r},t)$, $\textbf{A}(\mathbf{r},t)=\textbf{A}^{(0)}(\textbf{r},t)+\textbf{A}^{(p)}(\textbf{r},t)$.
For a photon scattering (one incoming and one outgoing photons) only
the term $\textbf{j}^{(1)}\textbf{A}^{(p)}(\textbf{r},t)$ of the interaction
Hamiltonian contributes, with
\begin{eqnarray}
\textbf{j}^{(1)}&=&\frac{\eta}{4\pi}\left\{\frac{\partial}{\partial t}\left[-2\textbf{B}^{(0)2}\textbf{E}^{(p)}+7(\textbf{E}^{(p)}\textbf{B}^{(0)})\textbf{B}^{(0)} \right]\right. \nonumber\\
&+&\left.2 \nabla\times \left[ 2(\textbf{B}^{(0)}(\textbf{B}^{(p)}\textbf{B}^{(0)})+\textbf{B}^{(0)2}\textbf{B}^{(p)}\right]\right\},
\end{eqnarray}
where $\textbf{j}^{(l)}$ denotes the term of the current
containing the probe field in the $l$-th order.
$\textbf{j}^{(2)}\textbf{A}^{(0)}(\textbf{r},t)$ can be responsible only
for processes with  a pair of photons in the initial or the final state.

First, we consider a photon scattering in a periodic field of a magnetic undulator of a linear polarization: $\textbf{B}^{(0)}(\textbf{r})=\textbf{B}_0 f(\textbf{r})$, with $f(\textbf{r})=\cos \textbf{q}\textbf{r}$. In this case,
\begin{eqnarray}
T_{21}=-2\pi i A_{k_1}^*A_{k_2}\delta (\omega_1-\omega_2)\int \left(\tilde{T}_{21} +\tilde{T}_{12}^*\right)e^{i(\textbf{k}_1-\textbf{k}_2)\textbf{r}}d^3\textbf{r},
\label{T21}
\end{eqnarray}
where $A_{k_i}=\sqrt{2\pi/\omega_i\mathcal{V}}$, $i=1,2$ and
\begin{eqnarray}
\tilde {T}_{21}&=&\frac{\eta\omega_1^2}{4\pi}\left\{ 7(\textbf{e}_1\textbf{B}_0)(\textbf{e}_2\textbf{B}_0)\right.\nonumber \\
&+&\left. 4 (\textbf{e}_1(\textbf{n}_1 \times \textbf{B}_0))\left[ (\textbf{e}_2(\textbf{n}_1 \times \textbf{B}_0))-i\left(\textbf{e}_2\left(\frac{\nabla f^2}{\omega_1f^2} \times \textbf{B}_0\right)\right)\right]\right.\nonumber \\
&+&\left. 2iB_0^2\left(\textbf{e}_2\left(\frac{\nabla f^2}{\omega_1f^2} \times \left(\textbf{n}_1 \times \textbf{e}_1\right)\right)\right)\right\}f^2(\textbf{r}),
\label{T21b}
\end{eqnarray}
where  $\textbf{n}_1=\textbf{k}_1/k_1$. The term $\tilde{T}_{21}$ corresponds to the case when the incoming photon is annihilated first, then the final photon is created, while in the cross-term $\tilde{T}_{12}^*$ this order is reversed.
The photon scattering probability per unit time $dW=n_1|T_{21}|^2d^3 \textbf{k}_2{\cal V}/(2\pi)^3$ reads:
\begin{eqnarray}
\frac{dW}{d\Omega}&=&\rho_1\omega^4|M_{21}+M_{12}^*|^2\mathcal{P},
\label{w}
\end{eqnarray}
with the number and the density of incoming photons $n_1$ and $\rho_1$, respectively, $\omega=k\equiv\omega_1=\omega_2$, the phase-matching factor
\begin{eqnarray}
\mathcal{P}=\left|\int_{({\cal V})} e^{{i\Delta \textbf{k} \textbf{r}}}d^3 \textbf{r}\right|^2,
\label{P_B}
\end{eqnarray}
and
\begin{eqnarray}
M_{21}&=&\frac{\eta}{16\pi}\left\{ 4((\textbf{n}_1\times \textbf{e}_1)\textbf{B}_0)((\textbf{n}_1\times \textbf{e}_2)\textbf{B}_0)\right. \nonumber\\
 &+& 7(\textbf{e}_1\textbf{B}_0)(\textbf{e}_2\textbf{B}_0)+\frac{8}{\omega}((\textbf{n}_1\times \textbf{e}_1)\textbf{B}_0)(\textbf{e}_2(\textbf{q} \times \textbf{B}_0)) \nonumber\\
&+& \left.\frac{4B_0^2}{\omega}(\textbf{e}_2( \textbf{q} \times (\textbf{n}_1\times \textbf{e}_1)))\right\}.
\label{M21}
\end{eqnarray}
The transition matrix element $M_{21}$ simplifies
when the incident photon momentum $\textbf{k}_1$ is perpendicular to $\textbf{B}_0$:
\begin{eqnarray}
M_{21}^{\bot}=\frac{7}{8\pi}\eta B_0^2,\,\,\,\, M_{21}^{||}=\frac{\eta B_0^2}{2\pi}\cos\vartheta,
\end{eqnarray}
where $M_{21}^{\bot}$ corresponds to the transverse  ($\textbf{B}_1 \bot \textbf{B}_0$) and $M_{21}^{||}$ to the longitudinal polarization ($\textbf{B}_1 || \textbf{B}_0$) and $\vartheta$ is the angle between $\textbf{k}_1$ and $\textbf{k}_2$.
The space integration in Eq. (\ref{P_B})
is carried out over the interaction volume ${\cal V}$, i.e., over the region of the overlap between the impinging laser beam with the undulator field, see the highlighted region in Fig. \ref{setup}(b). The factor $\mathcal{P}$ generates
the phase-matching Bragg condition $\Delta \textbf{k}=\textbf{k}_{2}-\textbf{k}_{1}-n\textbf{q}=0$ (with $n=2$ in the case of magnetic undulator) when the wave scattered from different spatial periods of the structure interfere constructively (Bragg interference). This can take place only at certain impinging angles of the probe wave:
\begin{eqnarray}
2k\sin\frac{\vartheta}{2}=nq,
\label{Bragg_condition}
\end{eqnarray}
The scattered radiation is concentrated in the specular direction.
Due to the Bragg interference, $\mathcal{P}\propto {\cal V}^2$ at exact phase-matching $\Delta \textbf{k}=0$  and the differential probability is proportional to ${\cal V}^2$. In particular, if the interaction region is rectangular
with corresponding lengths $L_x,L_y,L_z$,
\begin{eqnarray}
\mathcal{P}=\left( {\cal V}\,\,{\rm sinc}\frac{\Delta k_xL_x}{2}
 {\rm sinc}\frac{\Delta k_yL_y}{2} {\rm sinc}\frac{\Delta k_zL_z}{2}\right)^2 .
\label{sinc}
\end{eqnarray}
The total probability integrated over angular distribution is
\begin{eqnarray}
W= (2\pi)^2\rho_1 \omega^2|M_{21}|^2\int L_{z'}^2(\textbf{r}_{\bot})d^2\textbf{r}_{\bot},
\label{w_total}
\end{eqnarray}
where $L_{z'}(\textbf{r}_{\bot})$ is the length of the interaction region along the direction of $\textbf{k}_2$ and $\textbf{r}_{\bot}$ the coordinate transverse to $L_{z'}(\textbf{r}_{\bot})$ \cite{Landau}.
Estimating $\int L_{z'}^2(\textbf{r}_{\bot})d^2\textbf{r}_{\bot}\approx \bar{L}_{x'}\bar{L}_{y'}\bar{L}_{z'}^2$, with $\bar{L}_{z'},\,\bar{L}_{x'},\,\bar{L}_{y'}$ being the mean lengths of the interaction region along and transverse to $\textbf{k}_2$,
the number of scattered photons per unit time and volume is derived:
\begin{eqnarray}
\frac{d{\cal N}}{dtd\mathcal{V}}\approx {\cal G}\mathcal{W}(\vartheta)\alpha m^4\frac{B_0^4}{B_{cr}^4} \frac{E_0^2}{E_{cr}^2}\varepsilon,
\label{dN/dtdV}
\end{eqnarray}
where $E_0$ is the amplitude of the probe field, $ \mathcal{W}(\vartheta)\equiv |M_{21}|^2/(\eta B_0^2)^2$, ${\cal G}=(1/45)^2$,
$\varepsilon\equiv \bar{L}_{z'}/\lambda$ is the enhancement factor due to Bragg interference. The comparison of the Bragg scattering with the stimulated light-by-light scattering, proposed to realize with three strong laser beams \cite{Moulin,Lundstrom}, shows that the Bragg scattering can be larger by a factor $\bar{L}_{z'}/w_0$, with the laser beam waist size $w_0$,  which can amount to an order of magnitude. Note that Eq. (\ref{dN/dtdV}) is valid when the probe beam is rather monochromatic and has a low angular spread. The bandwidth ($\Delta \omega_1$) and the angular spread ($\Delta \vartheta$) of the probe beam should be limited to fulfill the Bragg condition Eq. (\ref{Bragg_condition}) within the energy uncertainty $\sim 1/\tau$, with the interaction time $\tau$:   $\Delta \omega_1 \lesssim (2\pi/\tau)D_{\omega_1}=(2\pi/\tau)(2/n^2)\sin^2(\vartheta/2)$ and  $\Delta \vartheta \lesssim (2\pi/\tau)D_{\vartheta}=(2\pi/\tau)(1/q\cos(\vartheta/2)$, where $D_{\omega_1 }=|\partial (\omega_1-\omega_2)/\partial \omega_1 |$ and $D_{\vartheta}=|\partial (\omega_1-\omega_2)/\partial \vartheta|$.
The latter impose restrictions on the enhancement factor $\varepsilon$. The largest interaction length $\bar{L}_{z'}$, equal to the undulator length $L_u$, is possible at $\vartheta \sim \pi$ when $k\sim q$. Then, the enhancement factor is determined by the number of undulator periods ($N_u$)  $\varepsilon \sim N_u$ but is restricted by the probe bandwidth $\varepsilon\lesssim \omega/\Delta \omega_1$.

Now let us consider Bragg scattering when the periodic structure is formed using a set of $N$ focused laser beams (elliptic Gaussian beams \cite{Yariv}) propagating parallel to each other:
%, see Fig. \ref{setup}(c):
\begin{eqnarray}
\textbf{E}^{(0)}=\textbf{E}_0 \sum_{n=1}^N f(x,y, z-nd)
\cos[\omega_L t-k_Lx+\varphi (x,y,z-n d)],
\end{eqnarray}
where $f(x,y, z)=( \sqrt{w_{y}w_{z}}/\sqrt{w_{y}(x)w_{z}(x)})e^{-y^2/w_{y}^2(x)+z^2/w_{z}(x)^2}$,  $\varphi(x,y,z)=-ky^2/2R_y(x)+z^2/2R_z(x)+\frac{1}{2}\tan^{-1}x/x_{0y}+\frac{1}{2}\tan^{-1}x/x_{0z}$, $w_{y}(x)=w_y\sqrt{1+x^2/x_{0y}^2}$, $w_{z}(x)=w_z\sqrt{1+x^2/x_{0z}^2}$, $R_y(x)=x+x_{0y}^2/x$, $R_z(x)=x+x_{0z}^2/x$, $x_{0y}=kw_y^2/2$ and $x_{0z}=kw_z^2/2$. The distance between the adjacent beams $d$ is assumed to be larger than the waist size of a single beam $w_{z}$, therefore, the superposition of fields of different beams is negligible. It is  also assumed that
$|\nabla f(\textbf{r}) \times \textbf{E}_0 |\ll k_Lf(\textbf{r})E_0$ and $|\nabla f(\textbf{r})|\ll k_LF(\textbf{r})$ and in this case $\textbf{B}_0=\textbf{n}_L\times \textbf{E}_0$, where $\textbf{n}_L=\textbf{k}_L/k$ and $\textbf{k}_L$ is the laser wavevector. The current calculated as
\begin{eqnarray}
\textbf{j}^{(1)}=\frac{\eta}{4\pi}\left(F^2(\textbf{r},t)\frac{\partial \textbf{R}_1}{\partial t} +F^2(\textbf{r},t)\nabla \times \textbf{R}_2 + \nabla F^2(\textbf{r},t)\times \textbf{R}_2\right),
\end{eqnarray}
where $\textbf{R}_1\equiv 4\left[(\textbf{E}^{(p)}\textbf{E}_0)-(\textbf{B}^{(p)}\textbf{B}_0) \right]\textbf{E}_0+7\left[(\textbf{E}^{(p)}\textbf{B}_0)+(\textbf{B}^{(p)}\textbf{E}_0)\right]\textbf{B}_0 $, $\textbf{R}_2\equiv \textbf{R}_1\times \textbf{n}_L$ and $F^2(\textbf{r},t)\equiv\sum_{l,m}^N f(x,y, z-ld)f(x,y, z-md)
\cos[\omega_L t-k_Lx+\varphi (x,y,z-l d)]\cos[\omega_L t-k_Lx+\varphi (x,y,z-m d)]$.
We consider only elastic scattering of photons. However, in this situation inelastic scattering with absorption or emission of additional laser photons can also take place: $\omega_1=\omega_2\pm2\omega_L$, which can be enhanced by the Bragg interference when coherence is met: $\textbf{k}_1=\textbf{k}_2\pm 2\textbf{k}_L+\textbf{q}$. In the case of elastic scattering, we average the current over the fast oscillations of nonresonant terms ($\cos^2(\omega_L t -k_Lx+\varphi)\rightarrow 1/2$). Then, the photon scattering probability in the setup of focused laser beams is given by Eq. (\ref{w}) with the following transition matrix element:
\begin{eqnarray}
M_{21}&=&\frac{\eta}{8\pi}\left\{4\left[(\textbf{e}_1\textbf{E}_0)+((\textbf{e}_1\times \textbf{n}_1)\textbf{B}_0)\right] \right.\nonumber\\
&\times& \left.\left[(\textbf{e}_2\textbf{E}_0)+\left(\textbf{e}_2\left(\left(\textbf{n}_1-n\frac{\textbf{q}}{k}\right)\times \textbf{B}_0\right)\right)\right]\right.\nonumber\\
&+& 7\left.\left[ (\textbf{e}_1\textbf{B}_0)-((\textbf{e}_1\times \textbf{n}_1)\textbf{E}_0)\right] \right.\\
&\times&\left.\left[(\textbf{e}_2\textbf{B}_0) -\left(\textbf{e}_2\left(\left(\textbf{n}_1-n\frac{\textbf{q}}{k}\right)\times \textbf{E}_0\right)\right)\right]\right\}.\,\, \nonumber
\end{eqnarray}
The phase-matching factor ${\cal P}$, after neglecting corrections proportional to the small diffraction parameter $kw_0\ll 1$, is:
\begin{eqnarray}
{\mathcal P}=\frac{(\pi L_xw_yw_z)^2}{16}e^{-\frac{\delta k_{y}^2w_y^2}{4}-\frac{\delta k_{z}^2w_z^2}{4}}
{\rm sinc}^2\frac{\delta k_xL_x}{2} \cdot\frac{\sin^2\frac{\delta k_zNd}{2}}{\sin^2\frac{\delta k_zd}{2}},
\label{PL}
\end{eqnarray}
where $\delta \textbf{k}=\textbf{k}_2-\textbf{k}_1$ and $L_x$ is the interaction length along the axis of the laser beams consisting the Bragg structure. The last multiplicative term in Eq. (\ref{PL}) has maxima at the Bragg condition $\delta k_z=nq$  with a bandwidth $\Delta(\delta k_z)\sim 2q/N$, where $q=2\pi/d$ and $n$ is an integer number. The exponential damping factor determines the maximal value for $\delta k_z\lesssim  2/w_z$, while the Bragg condition does the harmonic number $n\lesssim  d/\pi w_z$ (the damping factor is $e^{-n^2q^2w_z^2/4}$). 
\begin{figure}
\begin{center}
 \includegraphics[width=0.35\textwidth]{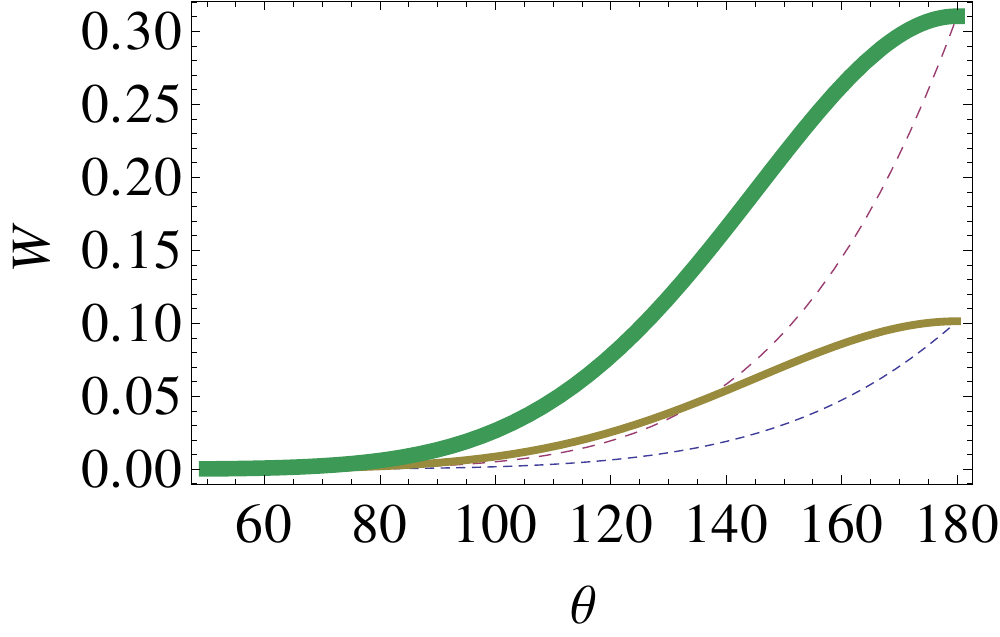}
 \caption{(Color online) The dependence of the scattering probability on the probe impinging angle [the factor $\mathcal{W}(\theta)$]: (solid line) B-field case, the upper curve - transverse polarization, the lower curve - longitudinal polarization; (dashed line) E-field case, the upper curve - longitudinal polarization, the lower curve - transverse polarization.} \label{theta}
\end{center}
\end{figure}
The strong laser field is linearly polarized. Two cases are possible: either $\textbf{B}_0$ (B-field) or $\textbf{E}_0$ (E-field) is perpendicular to the scattering plane formed by the wave vectors $\textbf{k}_1$ and $\textbf{k}_2$. The scattering matrix element for different polarizations of the probe wave reads:
\begin{eqnarray}
M_{21}^{B\parallel}&=&\frac{4\eta E_0^2}{\pi} \sin^4\frac{\vartheta}{4}\left(1-4\cos^2\frac{\vartheta}{4}\right).\nonumber\\
M_{21}^{B\bot}&=&\frac{7\eta E_0^2}{\pi}\sin^4\frac{\vartheta}{4}, \,\,\,\,
M_{21}^{E\parallel}=\frac{4\eta E_0^2}{\pi}\sin^4\frac{\vartheta}{4}\nonumber\\
M_{21}^{E\bot}&=&\frac{7\eta E_0^2}{\pi}\sin^4\frac{\vartheta}{4}\left(1-4\cos^2\frac{\vartheta}{4}\right),
\label{M_B||}
\end{eqnarray}
where the longitudinal (transverse) polarization means $B_1 \parallel (\bot)\,\, B_0$ in the case of B-field and $E_1 \parallel (\bot) \,\,E_0$ for E-field. 
The total number of scattered photons per unit volume and time in the laser beam setup is given by the same Eq. (\ref{dN/dtdV}) with $\mathcal{G}=e^{-n^2q^2w_z^2/4}(w_zq)^2/(\sqrt{2\pi}(360)^2)$ and replacing $B_{cr}\rightarrow E_{cr}$, $B_{0}\rightarrow E_{0}$. The angular dependence factor $\mathcal{W}(\theta)$ is shown in Fig. \ref{theta}. The enhancement factor in this case is
\begin{eqnarray}
\varepsilon&=& Nd/\lambda, \,\,\,\,\,\,\,\,\,{\rm if} \,\,\,\,\,kw_y^2/4\gtrsim Nd \nonumber \\
\varepsilon&=&2\pi\sqrt{Nd/\lambda}(w_y/\lambda), \,\,\,\,\,\,\,{\rm otherwise},
\label{enhancement}
\end{eqnarray}
with the wavelength $\lambda$. When $kw_y^2/4\lesssim Nd$, the emission angular width becomes restricted by the width of the energy uncertainty, as the Bragg condition should be fulfilled simultaneously.
We estimate the yield of scattered photons in the  setup of multiple laser beams. Employing a laser field with an intensity of $I=2.3\times 10^{22}$ W/cm$^2$, $\lambda=1 \mu$m, pulse duration of $\tau=100$ fs, $w_y\approx 3 \lambda$, $w_z\sim \lambda$, $d\approx \pi w_z$ and $N=10$, 
the Bragg scattering angle is close to $ \pi$  at $n=2$ resonance, and the number of scattered photons per pulse is
\begin{eqnarray}
{\cal N}=\frac{\alpha (2\pi)^{3/2}}{(360)^2{\rm e}^4}\left(\frac{I}{I_{cr}}\right)^3\frac{w_y^6c\tau}{\lambdabar_c^4\lambda^3}\left(\frac{\pi w_z}{d}\right)^2{\cal W}(\vartheta)\approx 4.8.
\label{estimation}
\end{eqnarray}
In the Bragg scattering setup, the enhancement factor $\varepsilon$ and the interaction volume are larger with respect to those in the stimulated light-by-light scattering, each roughly by an order of magnitude.  
The requirement for the vacuum background pressure to suppress the competing processes is the same as in the case of photon-photon scattering \cite{Lundstrom}.

Concluding, we have shown for the first time how the coherence effects arising in spatially structured
vacuum in strong periodic fields can enhance the vacuum polarization effects. In particular, the enhancement of the photon-photon scattering effect is proposed employing Bragg scattering of a probe laser beam by a set of parallel multiple laser beams. We want to underline that the considered coherence effect has a general nature. Similar enhancement effects will exist in all type of inelastic light-by-light scattering  and other processes based on spatially modulated vacuum polarization. The latter will be considered elsewhere.

We have benefited from fruitful discussions with C. H. Keitel and A. Di Piazza.
G. Yu. K. acknowledges the support and hospitality of the Max Planck Institute for Nuclear Physics. K. Z. H. acknowledges the hospitality of the Yerevan State University during a short visit.


\begin{thebibliography}{99}
\bibitem{Remark1} In fact, the $4$-momentum conservation in the constant magnetic field forces the 4-momentum of the outgoing photon $k_f$ to be equal to that of the incoming photon $k_i$. The latter takes place also in an  external laser field with a uniform amplitude because the two legs of the box related to the laser field describe absorption and emission of the laser photons. 
The absorption (emission) of two laser photons is prohibited by the energy-momentum conservation.
\end{thebibliography}
\end{document}